\documentclass[12pt]{article}

\usepackage[utf8]{inputenc}
\usepackage{amsfonts}
\usepackage{amsmath,commath}
\usepackage{amssymb}
\usepackage{booktabs}
\usepackage{cite}
\usepackage{xcolor}
\usepackage{colortbl}
\usepackage{enumitem}
\usepackage{epsfig}
\usepackage{epstopdf}
\usepackage{fancyhdr}
\usepackage{graphicx}
\usepackage{hyphenat}
\usepackage{IEEEtrantools}
\usepackage{ifthen}
\usepackage{mathtools}
\usepackage{multirow}
\usepackage{pdfpages}
\usepackage{rotating}
\usepackage{setspace}
\usepackage{subfigure}
\usepackage{textcomp}
\usepackage{url}
\usepackage{xfrac}
\usepackage{gensymb}
\usepackage{epigraph}
\usepackage{accents}
\usepackage{CJKutf8}
\usepackage{hyperref}
\usepackage{array}
\newcolumntype{P}[1]{>{\centering\arraybackslash}p{#1}}
\usepackage{hhline}

\usepackage{accents}

\newcolumntype{P}[1]{>{\centering\arraybackslash}p{#1}}
\usepackage{hhline}

\def\.{\cdot}

\def\##1{{\bf #1\mit}}
\def\_#1{{\bf #1\mit}}
\def\-#1{{\bf #1\mit}}
\def\=#1{\overline{\overline #1}}

\usepackage{caption}
\usepackage{sectsty}

\usepackage{listings}
\definecolor{mygreen}{RGB}{28,172,0} 
\definecolor{mylilas}{RGB}{170,55,241}

\def\e{\begin{equation}}
	\def\f{\end{equation}}
\def\_#1{{\bf #1}}
\def\.{\cdot}

\def\Re{{\rm Re\mit}}

\sectionfont{\fontsize{15}{20}\selectfont}
\subsectionfont{\fontsize{13}{15}\selectfont}

\begin{document}

\begin{center}
{\textbf{\large{Controlling  surface waves with temporal discontinuities of metasurfaces}}}\\
{X.~Wang,$^{1,2\ast}$~M.~S.~Mirmoosa,$^{1}$ and S.~A.~Tretyakov$^{1}$}\\
{$^1$Department of Electronics and Nanoengineering, Aalto University, Espoo, Finland}\\
{$^2$Institute of Nanotechnology, Karlsruhe Institute of Technology, Karlsruhe, Germany}\\
{$^{\ast}$Email:~xuchen.wang@kit.edu}
\end{center}
\vspace{1cm}


\begin{abstract}
In this paper, we investigate the scattering of surface waves on reactive impedance boundaries when the surface impedance undergoes a sudden change in time. We report three exotic wave phenomena. First, it is shown that by switching the value of the surface capacitance of the boundary, the velocity of surface waves can be fully controlled, and the power of reflected and transmitted surface waves are amplified.
Second, we show that when a capacitive boundary is switched to an inductive one, the surface wave stops completely, with a ``frozen'' static magnetic field distribution. The static magnetic fields are ``melt'' and restore propagating surface waves when the boundary is switched back to a capacitive one. 
Third, we show that temporal jumps of the boundary impedance couple free-space propagating waves to the surface wave, which is an analog to a spatial prism. These interesting effects enabled by temporal jumps of metasurface properties open up new possibilities for the generation and control of surface waves. 
\end{abstract}


\newpage


\section{Introduction}
The use of temporal discontinuities of material or surface parameters for controlling interactions of waves with electromagnetic or optical media has recently received significant attention. Temporal discontinuity/boundary means that the material property is spatially uniform but undergoes a sudden change in time, which is in analogy with a spatial interface where the material properties are discontinuous in space but uniform in time. 
The concept of the temporal boundary was introduced already in the middle of the last century \cite{morgenthaler1958velocity}, but later researchers mostly focused on temporal discontinuities of plasma and magnetoplasma media  \cite{kalluri2018electromagnetics}. 
Similarly to spatial boundaries, a temporal boundary can also generate reflection and transmission, but the scattering processes conserve momentum rather than energy, as at conventional spatial interfaces.
Due to the conservation of momentum, the frequency of wave should be converted at time discontinuities of material parameters. 
Such effects were experimentally confirmed at microwave and terahertz frequencies \cite{nishida2012experimental}, and recently in the optical range \cite{liu2021photon}. 
By cascading several temporal boundaries, it is also possible to eliminate reflections, in analogy with the quarter-wave transformer in microwave engineering \cite{pacheco2020antireflection, ramaccia2020light}. 

Recently, studies have been extended to more complex material properties, including dispersive media \cite{bakunov2021light, plansinis2015temporal},  anisotropic media \cite{akbarzadeh2018inverse, pacheco2020temporal, xu2022generalized}, chiral media \cite{mostafa2022temporal}, and magnetoplasma media \cite{li2022nonreciprocity}, revealing other  interesting wave phenomena, such as inverse prism \cite{akbarzadeh2018inverse}, temporal aiming \cite{pacheco2020antireflection}, nonreciprocity \cite{li2022nonreciprocity}, and polarization splitting \cite{mostafa2022temporal}. 
From the initial steps (scrutinizing effects of sudden changes of permittivity and permeability  \cite{morgenthaler1958velocity} and sudden creation of plasma \cite{wilks1988frequency} and magnetized plasma \cite{kalluri1996conversion}) till now, the research focus has been mainly on unbounded bulk media. 
There are several studies reporting effects at temporal discontinuities of other material structures, such as a leaky dielectric waveguide \cite{yin2022efficient} and a graphene sheet \cite{shirokova2019scattering, maslov2018temporal, wilson2018temporal}. 
However, temporal discontinuity of artificial two-dimensional (2D) materials, i.e., metasurfaces, has not been considered yet.

Metasurfaces or 2D metamaterials have shown strong potential due to the versatility of functionalities that they provide \cite{glybovski2016metasurfaces}. The electromagnetic properties of metasurfaces can be modeled by surface impedances, susceptibilities, and polarizabilities \cite{yang2019surface}.  
In this work, we investigate surface waves propagating on metasurface boundaries when the effective surface impedance changes rapidly from one value to another in time. We derive the reflection and transmission coefficients for surface waves after the temporal boundary, and find that the phase velocity of the surface wave can be freely controlled by switching the surface capacitance, realizing ultra-slow surface waves. We show also that the power carried by surface waves can be amplified after the jump. 
Moreover, when a reactive metasurface boundary is switched from  capacitive to inductive one, the surface wave will convert to a static magnetic field distribution over the boundary, and the magnetic stored energy is frozen. By switching back to a capacitive boundary, the electrical energy will be restored, and the surface wave starts to propagate again.
Finally, we show that a temporally discontinuous surface is able to couple a free-space propagating wave to a surface wave, acting as a temporal prism.


\section{Dispersion diagram of time-invariant reactive boundaries}

\begin{figure}[t]
\centering \includegraphics[width=0.6\linewidth]{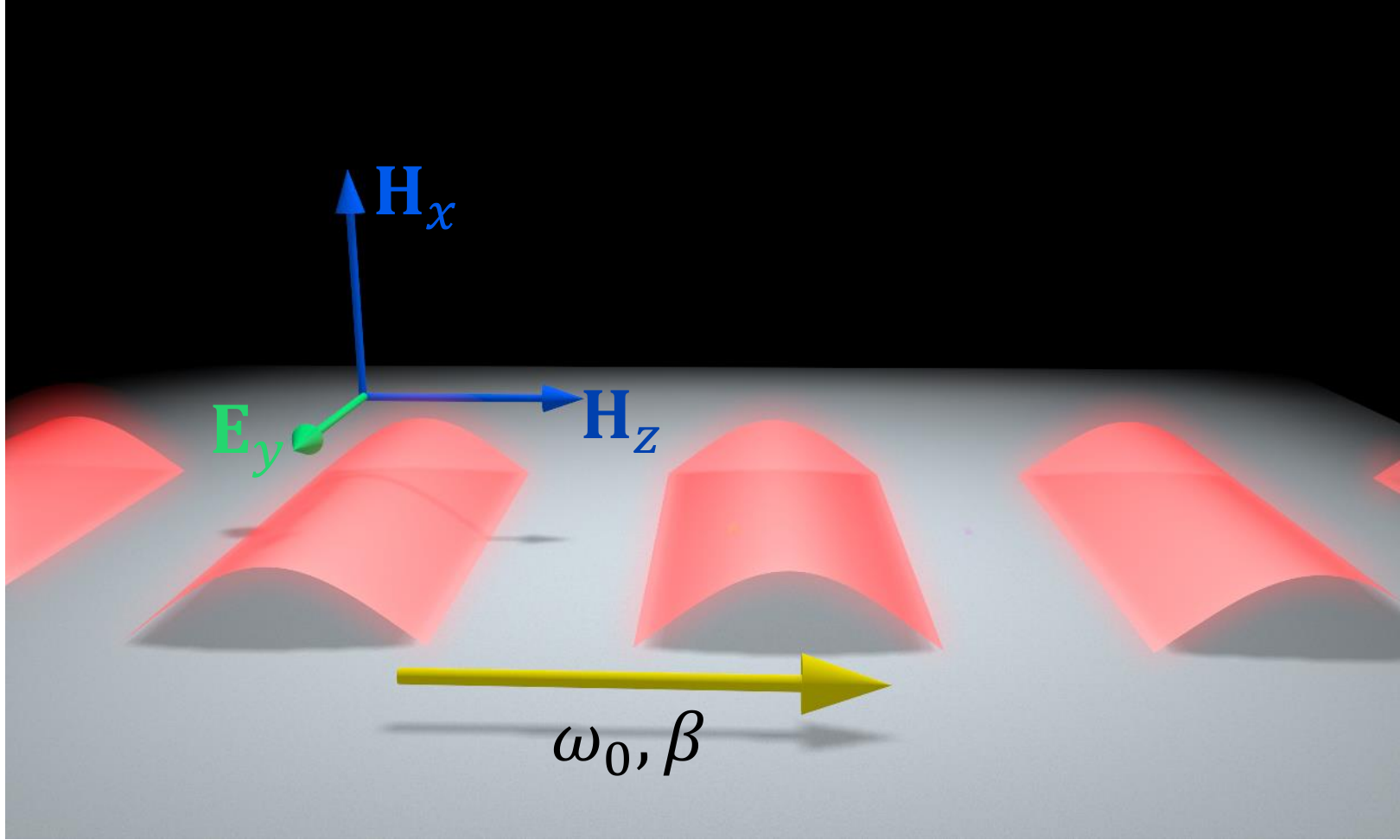}
\caption{A TE-polarized surface wave which is propagating along an impenetrable capacitive boundary.} 
\label{fig:schematics}
\end{figure}

\begin{figure}[t]
\centering \includegraphics[width=0.6\linewidth]{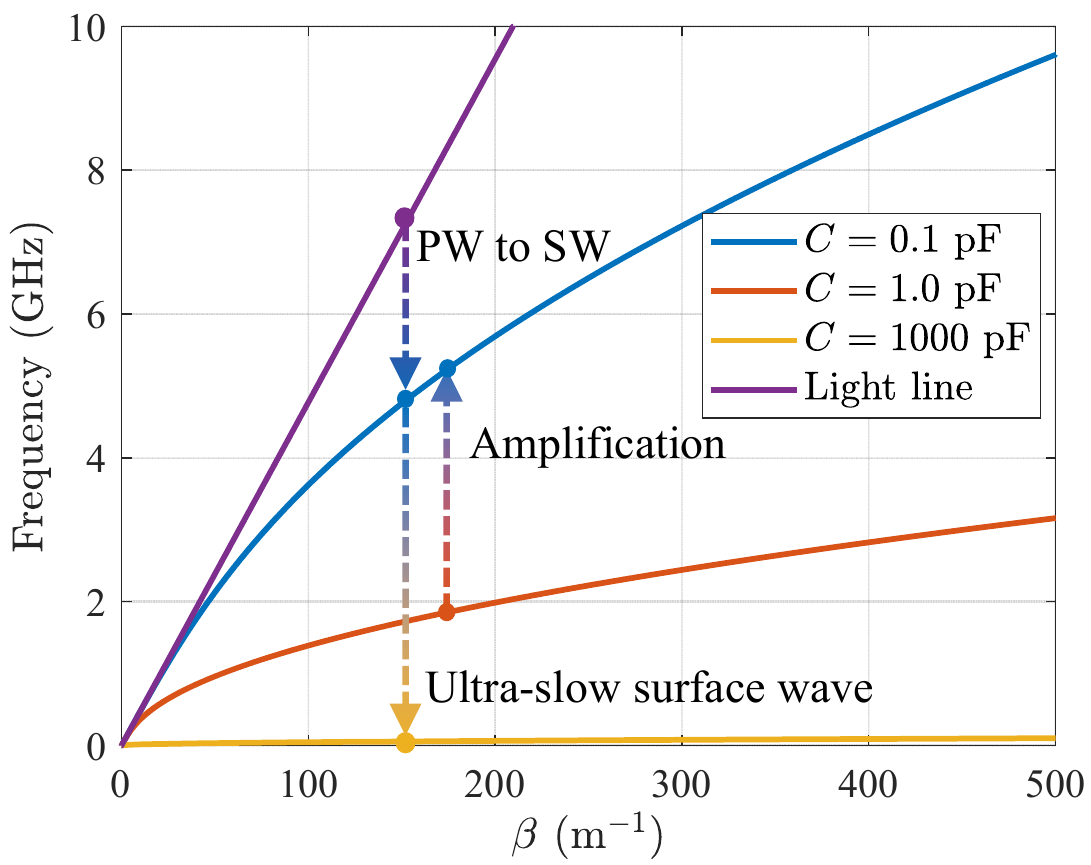}
\caption{Dispersion relation of a capacitive boundary. PW and SW are abbreviations of plane wave and surface wave, respectively.} \label{fig:dispersion}
\end{figure}

Let us consider an impenetrable metasurface which is characterized by a capacitive surface impedance with the surface capacitance $C$. First, we remind the well-known dispersion relation for waves traveling along such impedance boundaries in the stationary case. Since the surface impedance is capacitive, transverse electric (TE)-polarized waves can propagate along the surface, as shown in Fig.~\ref{fig:schematics}. We write the electric field as $\mathbf{E}_y=E_0e^{-j\beta z-\alpha x}e^{j \omega t}\mathbf{y}$, where $\beta$ is the propagation constant along the surface, $\alpha$ is the decay factor along the $x$-direction, and $E_0$ represents the wave amplitude. The field corresponding to the surface wave exists in free space. Hence, it must satisfy the Helmholtz equation written for free space, i.e., $\nabla^2\mathbf{E}_y-\mu_0\epsilon_0 {\partial^2\mathbf{E}_y}/{\partial t^2}=0$. By substituting the electric field in the Helmholtz equation, we arrive at the relation $\alpha^2-\beta^2+\omega^2\epsilon_0\mu_0=0$. In addition, we also substitute the electric field in Faraday's law $\nabla\times\mathbf{E}=-\mu_0{\partial \mathbf{H}}/{\partial t}$, and, as a result, we calculate the tangential component of the magnetic field associated with the surface wave: $\mathbf{H}_z=\frac{\alpha E_0}{j\omega \mu_0}e^{-j\beta z-\alpha x}e^{j \omega t}\mathbf{z}$ (note that TE-polarized waves have two components of magnetic field). Now, the tangential components of the fields at the boundary $x=0$ respect the boundary condition which is ${j\omega C}\cdot\mathbf{E}_y=\mathbf{n}\times \mathbf{H}_z$. Imposing this condition gives us the dispersion relation
\begin{equation}
\beta^2=k_0^2\Big(1+\eta_0^2\omega^2C_0^2\Big), 
\end{equation} 
in which $k_0=\omega\sqrt{\epsilon_0\mu_0}$ and $\eta_0=\sqrt{\mu_0/\epsilon_0}$ are the free-space wavenumber and intrinsic impedance, respectively. Figure~\ref{fig:dispersion} presents the dispersion curves for different boundaries characterized by different values of surface capacitances. As the figure explicitly shows, if the value of $C$ decreases, the curve shifts towards the light line, and if the value of $C$ grows, the curve shifts towards the horizontal axis (the propagation constant axis). In the following, we explain how we can efficiently use this property together with temporal discontinuities in order to engineer the phase and energy velocity of surface waves.  

\section{Forward and backward surface waves}

Let us assume that at some moment of time  ($t=0$)  the surface capacitance quickly changes  from one value to another one. This temporal jump generates forward and backward surface waves. In fact, this is in analogy with temporal jumps of volumetric material parameters such as permittivity and permeability. In this latter case, we know that since the material is uniform in space, the phase constant (or the propagation constant) is conserved meaning that it does not change at the moment of temporal discontinuities of material parameters. However, the angular frequency changes. Interestingly, the equation for the new angular frequency has two solutions whose absolute values are the same but they are different in sign. The positive value corresponds to a forward wave, and the negative value is related to a backward wave (there is an analogy with time reversal). Similarly, for our specific scenario, at a time jump of surface impedance, the propagation constant  $\beta$ along the surface is fixed, because the surface is assumed to be spatially uniform at all times. Thus, after the jump, we have two surface waves propagating over the surface in opposite directions. As the next step, we will find the amplitudes of these two waves. 

We suppose  that before the moment of capacitance jump, a TE surface wave (propagation constant $\beta$, frequency $\omega_0$) is propagating over a  capacitive boundary, with electric and magnetic fields
\begin{equation}
    \mathbf{E}_{\rm i}=E_0e^{-j\beta z-\alpha_0  x}e^{j \omega_0 t}\mathbf{y}, \quad
     \mathbf{H}_{\rm i}=\frac{E_0\alpha_0}{j\omega_0 \mu_0} e^{-j\beta z-\alpha_0  x}e^{j \omega_0 t}\mathbf{z}-
   \frac{E_0\beta}{\omega_0 \mu_0} e^{-j\beta z-\alpha_0  x}e^{j \omega_0 t}\mathbf{x}. \label{eq: incident}
\end{equation}
We call this wave \emph{incident}.
At $t=0$, the surface capacitance suddenly changes from $C_0$ to $C_1$. In order to adapt to the new boundary, the wave frequency should be converted to $\omega_1$, according to the dispersion relation. Thus, we have (after a short transient period) 
\begin{equation}
 \mathbf{E}_{\rm r}=R_{\rm E}E_0e^{-j\beta z-\alpha_2 x}e^{-j \omega_1 t}\mathbf{y},\quad \mathbf{E}_{\rm t}=T_{\rm E}E_0e^{-j\beta z-\alpha_1 x}e^{j \omega_1 t}\mathbf{y},
\label{eq:EFFB}
\end{equation}
where $R_{\rm E}$ and $T_{\rm E}$ define the amplitudes of the backward and forward waves, respectively. We call them \emph{reflection and transmission coefficients} of electric field. Note that in the above equation, as explained before, the propagation constant is conserved and the angular frequency changes. This property gives rise to a different value for the attenuation constant of the fields above the boundary (parameter $\alpha$). Having the expressions of the electric field, the magnetic fields of the incident, reflected, and transmitted fields are obtained by employing Maxwell's equations, leading to 
\begin{subequations}
\begin{align}
         \mathbf{H}_{\rm r}&=\frac{-R_{\rm E}E_0\alpha_1}{j\omega_1 \mu_0} e^{-j\beta z-\alpha_1  x}e^{-j \omega_1 t}\mathbf{z}+
   \frac{R_{\rm E}E_0\beta}{\omega_1 \mu_0} e^{-j\beta z-\alpha_1  x}e^{-j \omega_1 t}\mathbf{x},\\[8pt]
   \mathbf{H}_{\rm t}&=\frac{T_{\rm E}E_0\alpha_1}{j\omega_1 \mu_0} e^{-j\beta z-\alpha_1  x}e^{j \omega_1 t}\mathbf{z}-
   \frac{T_{\rm E}E_0\beta}{\omega_1 \mu_0} e^{-j\beta z-\alpha_1  x}e^{j \omega_1 t}\mathbf{x}. 
\end{align} 
\label{eq: magnetic fields}
\end{subequations}

Regarding Maxwell's equations, we have the time derivative of the electric and magnetic flux densities, and, in accordance with this fact, these vectors should be continuous at the moments of abrupt changes of the surface impedance. Therefore, \begin{equation}
\mathbf{D}_{t=0^-}=\mathbf{D}_{t=0^+}, \quad \mathbf{B}_{t=0^-}=\mathbf{B}_{t=0^+}.\label{eq:condition}
\end{equation} 
However, since the environment allowing surface-wave propagation is fixed (free space), the continuity described by the above equation results in the continuity of the electric and magnetic fields. Let us choose one point infinitesimally close to the boundary,  at $x=\delta$, where $\delta\rightarrow 0^+$. First, based on the electric field continuity, $\mathbf{E}_{\rm i}|_{t=0^-}=(\mathbf{E}_{\rm r}+\mathbf{E}_{\rm t})|_{t=0^+}$, it is easy to show that $T_{\rm E}+R_{\rm E}=1$. Second, according to the continuity of the magnetic field at the chosen point, $\mathbf{H}_{\rm i}|_{t=0^-}=(\mathbf{H}_{\rm r}+\mathbf{H}_{\rm t})|_{t=0^+}$, we can also show that $T_{\rm E}-R_{\rm E}=\omega_1/\omega_0$. In fact, this identity is achieved by considering only the $x$ component of the magnetic field. We do not use the $z$ component of the magnetic field because it is derived based on the derivative of the electric field of the surface wave with respect to $x$. But, in this direction, there is a discontinuity in space. However, in contrast, $\mathbf{H}_{ x}$ is obtained from the derivative of the electric field with respect to $z$, and, in this direction, there is no spatial discontinuity. Therefore, only $\mathbf{H}_x$ is required to be continuous. Finally, the reflection and transmission coefficients related to the electric fields are calculated as 
\begin{equation}
T_{\rm E}=\frac{1}{2}\left(1+\frac{\omega_1}{\omega_0}\right), \quad R_{\rm E}=\frac{1}{2}\left(1-\frac{\omega_1}{\omega_0}\right).
\label{eq: TR}
\end{equation}
Notice that based on Eqs.~\eqref{eq:EFFB} and \eqref{eq: magnetic fields}, we can also define reflection and transmission coefficients in terms of the $x$-component of the magnetic field, i.e., $\mathbf{H}_{{\rm r}x}=R_{\rm  H}\mathbf{H}_{{\rm i}x}$ and $\mathbf{H}_{{\rm t}x}=T_{\rm  H}\mathbf{H}_{{\rm i}x}$. According to this definition, we obtain that 
\begin{equation} 
T_{\rm H}=\frac{1}{2}\left(1+\frac{\omega_0}{\omega_1}\right), \quad R_{\rm H}=\frac{1}{2}\left(1-\frac{\omega_0}{\omega_1}\right).\label{eq: TR}
\end{equation}



\section{Examples of phenomena at time jumps of boundary properties}
Next, we provide four examples of important surface-wave phenomena on  temporal boundaries, demonstrating freezing, amplification,  and excitation of surface modes. 

\subsection{Surface-wave amplification}
At a temporally varying interface,  the passivity condition does not necessarily hold since energy may enter or leave the system through the external device that changes the surface parameters.
Here, we examine gain of the system that is defined  as the ratio of incident and the scattered powers (power density integrated from $x=0$ to $x=+\infty$), i.e., $G={(|P_{\rm r}|+|P_{\rm t}|)}/{|P_{\rm i}|}$. Note that the unit of power is W/m here.
By substituting the electric and magnetic fields, the gain parameter can be evaluated as
\begin{equation}
    G=\frac{\int_0^{+\infty} |\Re(\mathbf{E}_{\rm r}\times \mathbf{H}_{\rm r}^*)| {dx} +\int_0^{+\infty}|\Re(\mathbf{E}_{\rm t}\times \mathbf{H}_{\rm t}^*)|dx}{\int_0^{+\infty}|\Re(\mathbf{E}_{\rm i}\times \mathbf{H}_{\rm i}^*)|dx}=\frac{\alpha_0}{2\alpha_1}\left(\frac{\omega_0}{\omega_1}+\frac{\omega_1}{\omega_0}\right). \label{eq: gain}
\end{equation}
Clearly, gain is always larger than unity if $\alpha_0>
\alpha_1$, which is  equivalent to  $\omega_1>\omega_0$ or $C_0>C_1$. The sharper is the jump, the more energy will be injected into the system. If  the capacitance does not change ($\omega_0=\omega_1$ and $\alpha_0=\alpha_1$) at $t=0$, then, obviously, $G=1$.

\begin{figure}[h]
\centering \includegraphics[width=0.8\linewidth]{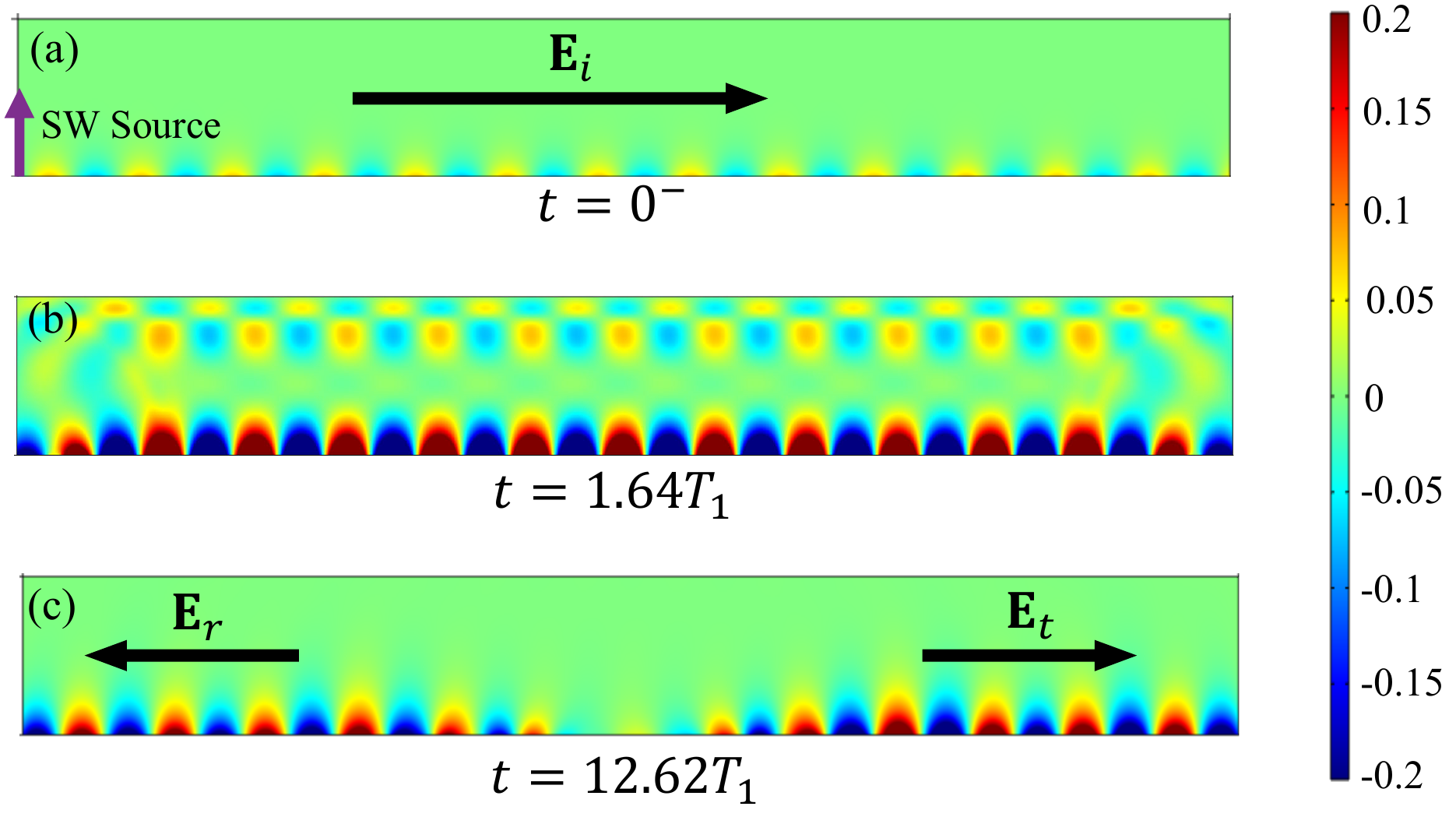}
\caption{Electric fields distribution at (a) $t=0^-$,   (b) $t=1.64T_1$, and (c) $t=12.62T_1$. } \label{fig:field_amplfication}
\end{figure}

\begin{figure}[h]
\centering \includegraphics[width=1\linewidth]{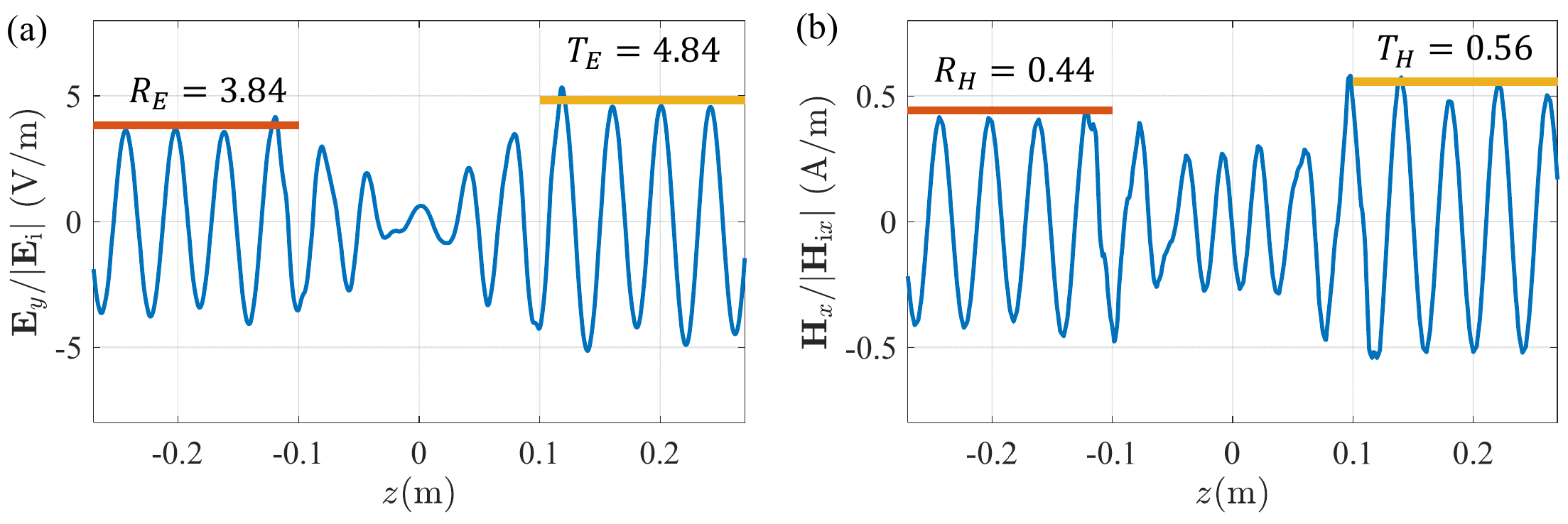}
\caption{Normalized (a) electric and (b) magnetic  fields  variations along the surface. The red and orange lines show the theoretical values of transmission and reflection coefficients.} \label{fig:TR_amplfication}
 \end{figure}

Next, we numerically verify the reflection, transmission, and amplification phenomena.  
As an example, we assume that before the jump $C_0=10$~pF and after the jump $C_1=0.1$~pF. The eigenfrequencies at these two states are calculated as $\omega_0/(2\pi)=0.556$~GHz and $\omega_1/(2\pi)=4.82$~GHz, with the same propagation constant along the surface $\beta=153$~m$^{-1}$. Figure~\ref{fig:field_amplfication} shows the fields simulated in COMSOL. The surface wave was launched from the left boundary of the computation domain, and it propagates to the right  boundary. At $t=0^-$, the surface wave is uniformly distributed over the boundary with the surface capacitance  $C_0$ [see Fig.~\ref{fig:field_amplfication}(a)]. At $t=0$ the excitation source is removed, and the boundary capacitance is switched to $C_1$. While the new surface eigenmode is established at the changed boundary, transient  scattering is generated, as shown in Fig.~\ref{fig:field_amplfication}(b). The transmitted and reflected surface waves form a standing wave.
At $t=12.62~T_1$, the two oppositely traveling surface waves are fully separated. One can explicitly see in Fig.~\ref{fig:field_amplfication}(c) that the electric fields of the transmitted and reflected surface waves are much stronger than the incident wave in Fig.~\ref{fig:field_amplfication}(a).
The calculated electric fields at $t=12.62T_1$ on the boundary are shown in Fig.~\ref{fig:TR_amplfication}(a), which agrees well with the theoretically predicted values. The magnetic fields (the $x$-component) of the two waves reduce to around half of the incident wave amplitude. By summing up the powers of two reversely propagating surface waves, the gain of the system can be confirmed as $G=4.32$.

This conclusion of surface wave amplification is in contrast to that in Ref.~\cite{maslov2018temporal}, where the authors state that  wave amplification is not possible at a temporal interface of a graphene sheet.

\subsection{Ultra-slow surface waves}

Next, we consider the scenario of switching surface capacitance from a low to a very high value ($C_0$ to $C_1$, with $C_1\gg C_0$). 
Figure~\ref{fig:dispersion} shows that when the sheet capacitance $C_1$ increases, the eigenfrequency decreases, corresponding to a  reduced phase velocity $\omega_1/\beta$. When $C_1\rightarrow +\infty$, the phase velocity approaches  zero. 
More interestingly,  the {\it group velocity} $d\omega/d\beta$ also tends to zero due to the extreme flatness of the dispersion curve. This method of wave freezing is very different from the conventional method for slowing down or freezing light by engineering the dispersion curve of photonic structures, where only zero group velocity is achieved. 

\begin{figure}[h]
\centering \includegraphics[width=0.85\linewidth]{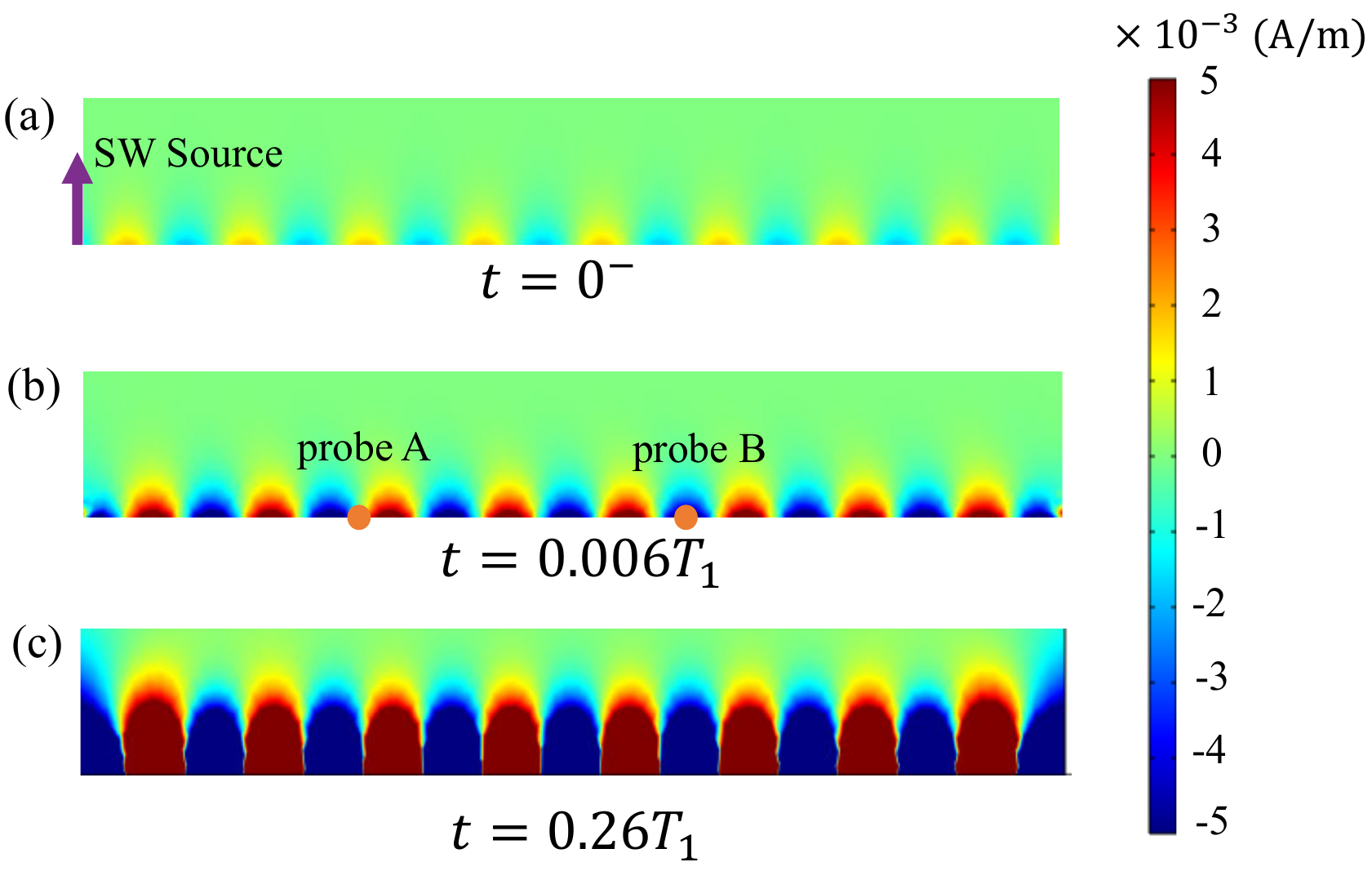}
\caption{Spatial distributions of magnetic fields $\mathbf{H}_x$ at (a) $t=0^-$,   (b) $t=0.006T_1$, and (c) $t=0.26T_1$. } \label{fig:fields_slow}
\end{figure}

\begin{figure}[h]
\centering \includegraphics[width=0.8\linewidth]{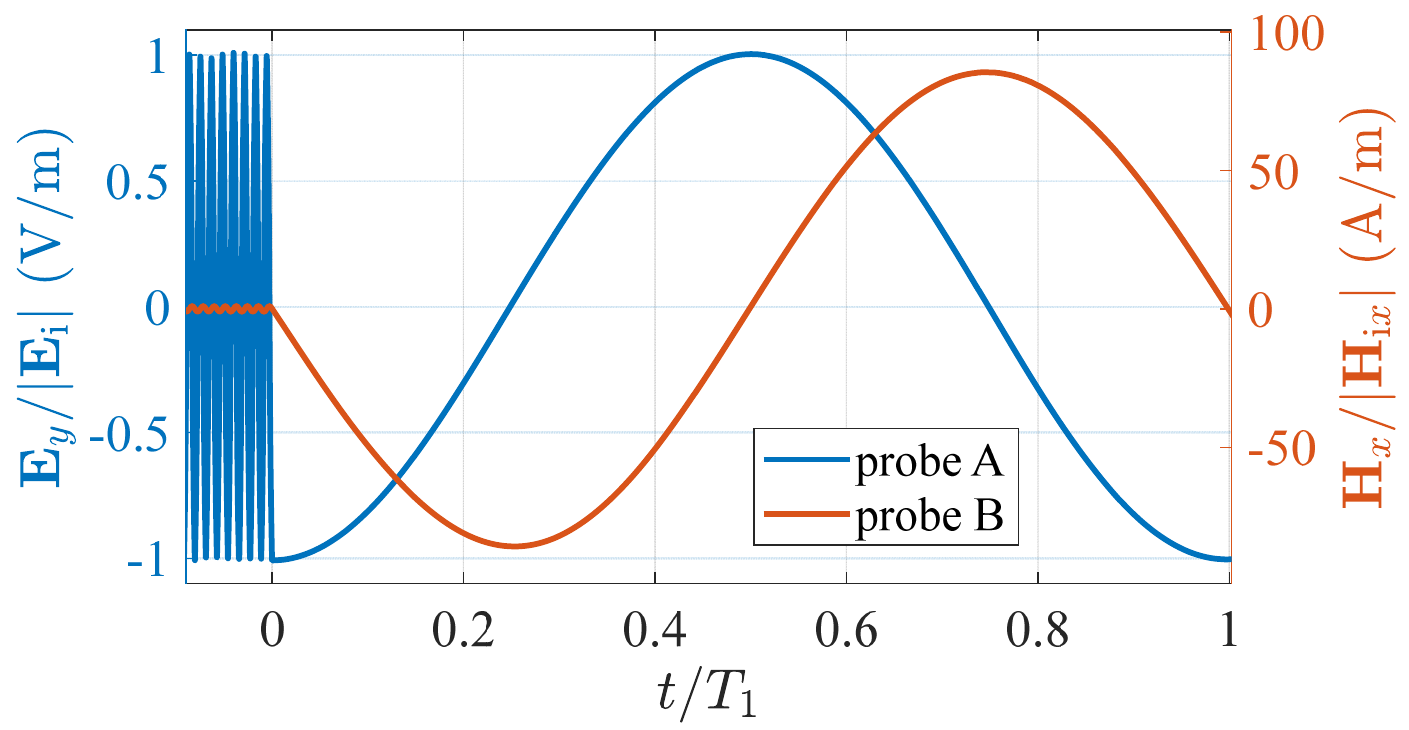}
\caption{Temporal variations of $\mathbf{E}_y$ at probe A and  $\mathbf{H}_x$ at probe B.   The fields are normalized by their incident amplitudes. } \label{fig:TR_slow}
\end{figure}

In numerical examples, we assume $C_0=0.1$~pF,  $C_1=1000$~pF, and  $\beta=153~\rm m^{-1}$. The corresponding eigenfrequencies are $\omega_0/(2\pi)=4.83~\rm GHz$ and $\omega_1/(2\pi)=0.055~\rm GHz$. Figure~\ref{fig:fields_slow} shows the distributions of $\mathbf{H}_x$ in the simulation space  before and after the temporal jump. The magnetic field amplitude dramatically increases after the jump. This is the reason why the total energy of the system increases with a significant gain ($G=44$).
We position two probes [probe A and probe B in Fig.~\ref{fig:fields_slow}(b)] on the boundary. Probes A and B are placed at the antinodes (maxima) of $\mathbf{E}_y$ and $\mathbf{H}_x$, respectively. 
The time-varying electric field at probe A and magnetic field at probe B are recorded and shown in Fig.~\ref{fig:TR_slow}. 
Noticeably,  the oscillation frequency of fields after the jump ($t>0$) is significantly reduced, while the propagation constant $\beta$ remains unchanged, as seen in Fig.~\ref{fig:fields_slow}. 
This result clearly demonstrates that the reflected and transmitted waves are ultra-slow. 
According to Eq.~\eqref{eq: TR}, the electric field amplitude of the reflected and transmitted waves can be estimated as  $T_{\rm E}={1}/{2}$ and $R_{\rm E}=-{1}/{2}$. Thus, a standing wave is formed, with the amplitude $|T_{\rm E}|+|R_{\rm E}|=1$ at the field maximum, which is confirmed in Fig.~\ref{fig:TR_slow}.
It is worth mentioning that waves cannot be completely stopped by this technique because in this case the energy becomes infinite, as predicted by Eq.~\ref{eq: gain}. 

\subsection{Frozen surface waves}

\begin{figure}[h]
\centering \includegraphics[width=0.75\linewidth]{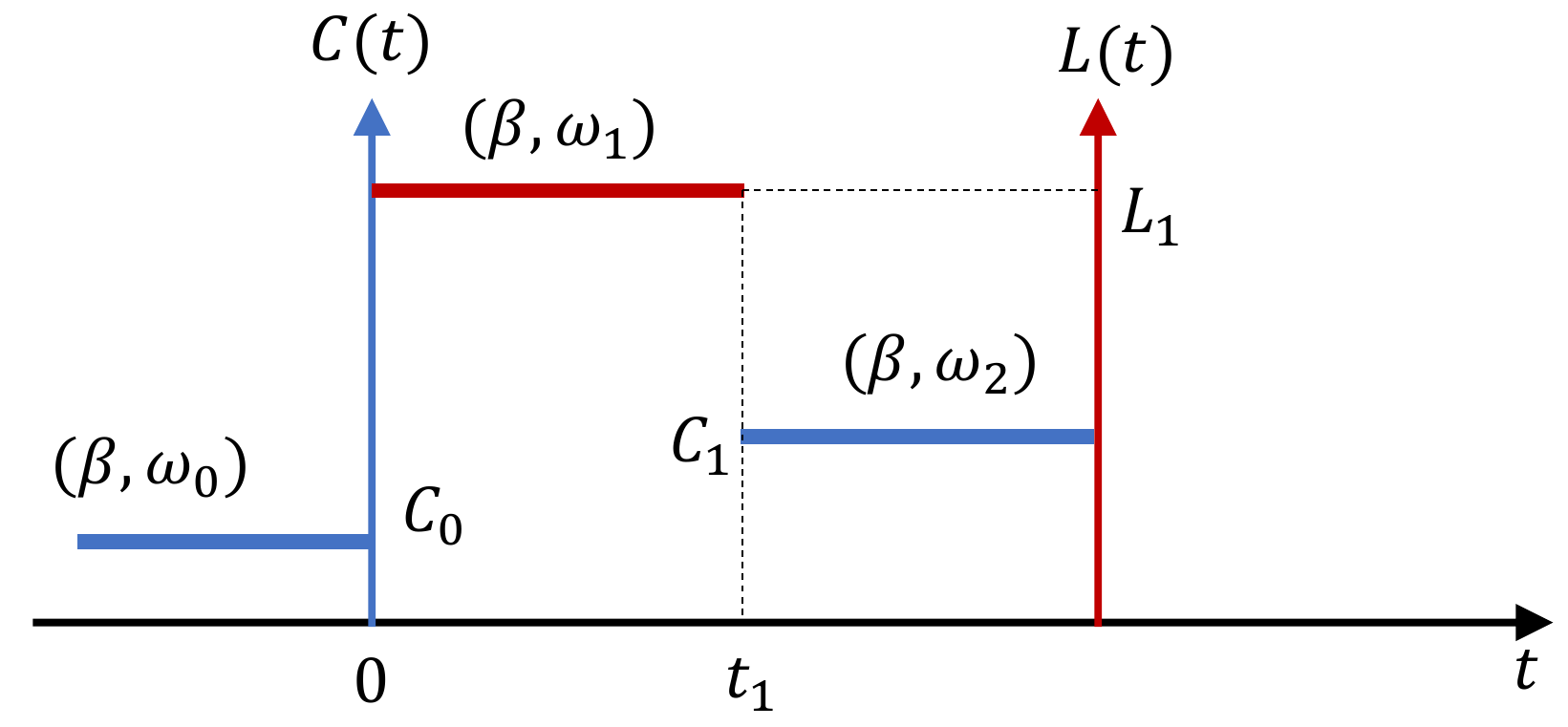}
\caption{Boundary property as a function of time. Here, $C_0=1$~pF, $L_1=0.1$~nH, $C_1=0.1$~pF, $\beta=153$~(m$^{-1}$), $\omega_0=1.73$~GHz, $\omega_1=0$~GHz, and $\omega_2=4.82$~GHz. } \label{fig:jump}
\end{figure}

In this part of section, we investigate how surface wave behaves when the sheet reactance changes from capacitive to inductive, and then it returns back to capacitive ($C_0 \rightarrow L_1 \rightarrow C_1$), as shown in Fig.~\ref{fig:jump}. We will show that after the first jump, the surface wave stops, and the electrical fields disappear, while  non-uniform static magnetic fields remain frozen over the boundary. The electric field is re-created after the second jump, and the surface wave again starts to propagate.

\begin{figure}[h]
\centering \includegraphics[width=0.8\linewidth]{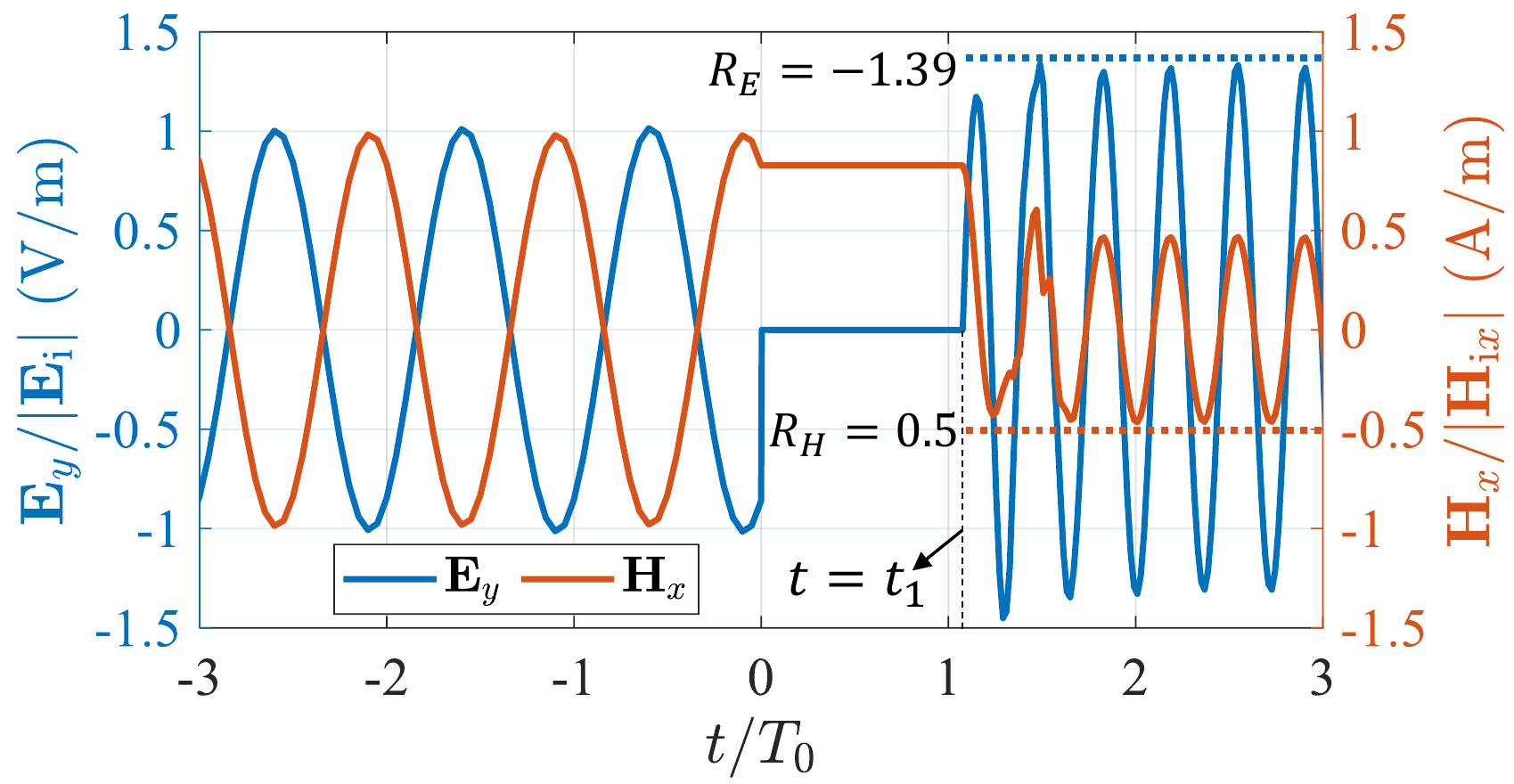}
\caption{Temporal variations of $\mathbf{E}_y$ and $\mathbf{H}_x$ at the probing point near the excitation side.   The fields are normalized by their incident amplitudes. } \label{fig:frozen}
\end{figure}

Let us consider a capacitive boundary $C_0$ over which a TE surface wave ($\beta, \omega_0$) propagates. The expressions for the incident electric and magnetic fields are found in Eq.~\ref{eq: incident}. 
At $t=0$, the boundary is suddenly changed to an inductive one, with the surface inductance $L_1$, and  the excitation is removed. Unequivocally, the existing TE surface mode cannot be supported by the inductive boundary which only supports the TM surface mode. However, this statement holds only for an oscillating TE surface mode at a non-zero frequency. We notice that, if the frequency of the TE mode is converted to zero ($\omega_1=0$) after the jump, such static TE mode can satisfy the inductive boundary condition as well as the Maxwell's equations in free space. 
Since the jump must conserve the $x$-component of magnetic field, we can write this component after the jump as $\mathbf{H}_x^{t=0^+}=H_0e^{-j\beta z-\alpha_1 x}\mathbf{x}$, where $H_0=-\frac{E_0\beta}{\omega_0 \mu_0}$. Note that the time-harmonic term $e^{j\omega_1t}$ disappears because $\omega_1=0$. Using the Maxwell's equations,
the other field components can be determined as $\mathbf{H}_y^{t=0^+}=0$, $\mathbf{H}_z^{t=0^+}=jH_0e^{-j\beta z-\alpha_1  x}\mathbf{z}$, and $\mathbf{E}_{x}^{t=0^+}=\mathbf{E}_{y}^{t=0^+}=\mathbf{E}_{z}^{t=0^+}=0$.
One can see that the electric field vanishes, and there is static magnetic field which  is spatially non-uniform both in $z$ and $x$ directions.
Figure~\ref{fig:frozen_field_distribution} shows the static field distribution in space when the boundary parameter is switched from $C_0=1$~pF to $L_1=0.1$~nH. 
Since the static magnetic field satisfies the boundary condition of the inductive boundary  $j\omega_1 L_1\cdot(\mathbf{n}\times \mathbf{H}_z^{t=0^+})=\mathbf{E}_y^{t=0^+}=0$, it can be regarded as the ``frozen eigenmode'' of the inductive boundary.  Figure~\ref{fig:frozen} shows the instantaneous field variation at the probing point near the illuminating side (the probe position is shown in Fig.~\ref{fig:frozen_field_distribution}). 
One can see that the magnetic field ($\mathbf{H}_x$ and $\mathbf{H}_z$) stops to oscillate and remains constant in time  after the boundary capacitance jumps to inductance and the electric field reduces to zero.
\textbf{\begin{figure}[h]
\centering \includegraphics[width=0.85\linewidth]{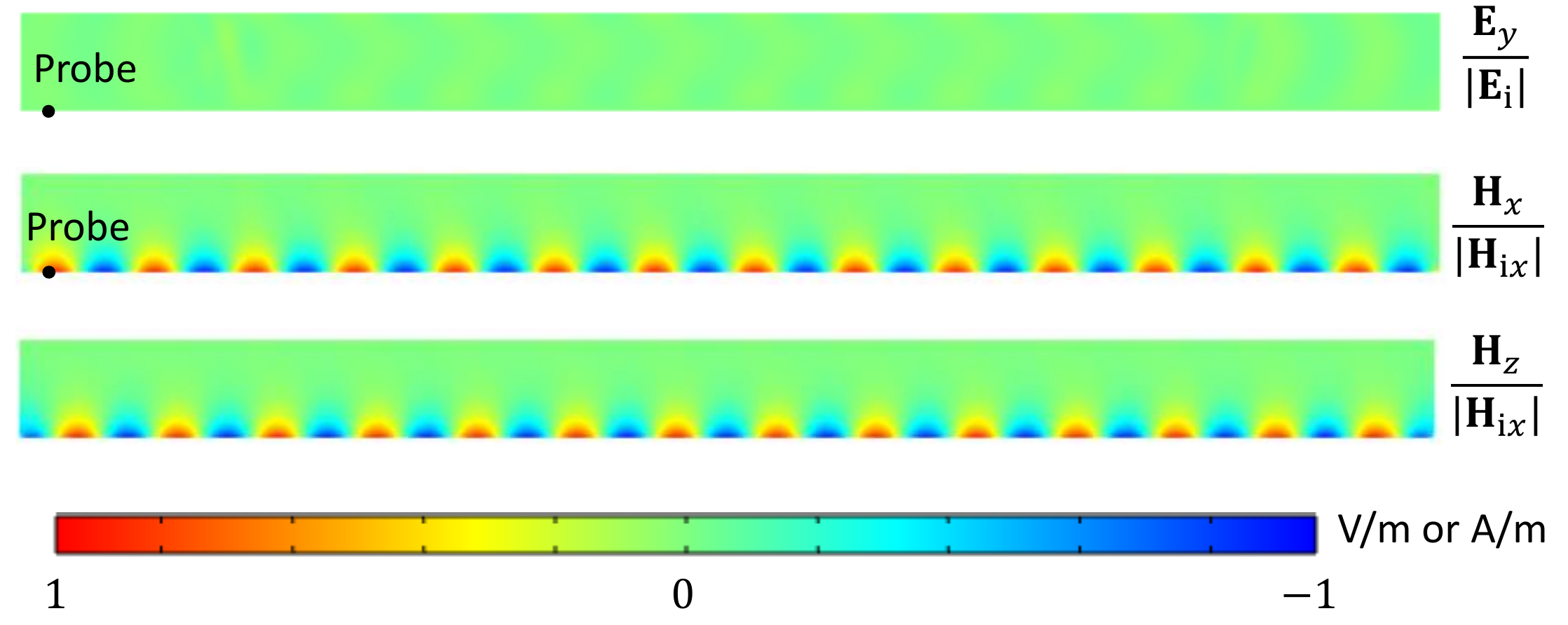}
\caption{Spatial distributions of the static electric and magnetic fields at the freezing state. } \label{fig:frozen_field_distribution}
\end{figure}}


At $t=t_1=1.076~T_0$, where $T_0$ is the temporal period of the incident wave, the inductive boundary is switched back to a capacitive one with the surface capacitance $C_1=0.1$~pF. The frequency of the surface mode must be converted to $\omega _2$ in order to satisfy the eigenvalue equation of the new capacitive boundary ($\beta, \omega_2$). Therefore, the vanished electric field revives after the second jump. By using the temporal boundary condition of Eq.~\ref{eq:condition}, we can calculate the reflection and transmission coefficients after the second jump: 
\begin{equation}
    T_{\rm E}=-R_{\rm E}=\frac{\omega_2}{2\omega_0}, \quad T_{\rm H}=R_{\rm H}=\frac{1}{2},
\end{equation}
where the coefficients are normalized by the incident amplitudes.
The above equation indicates that the amplitude of the total electrical field on the boundary can be amplified if $\omega_2>\omega_0$, or equivalently $C_1<C_0$. The normal component of magnetic field of the reflected and transmitted waves is always equal to half of the incident wave amplitude, independently of the value of $C_1$. As shown in Fig.~\ref{fig:frozen}, the amplitude of the reflected mode is very close to the theoretical predictions (marked as dashed lines).  

It is worth mentioning that creation of a static magnetic field distribution at a temporal boundary was noticed in early works on wave propagation in a suddenly created plasma \cite{zero_1,zero_2,wilks1988frequency} and in magnetized plasma with quickly changed bias field \cite{kalluri1996conversion,li2022nonreciprocity}. This mode was called \emph{wiggler mode}.  
Here, in studying surface waves, we have found that a traveling wave can be completely converted into a static mode, that can survive independently. Furthermore, at a later time, it can be used for restoring the propagating wave when the second jump is made. This effect can be used to temporarily stop a light pulse. 

\subsection{Temporal prism}

\begin{figure}[h]
\centering \includegraphics[width=0.8\linewidth]{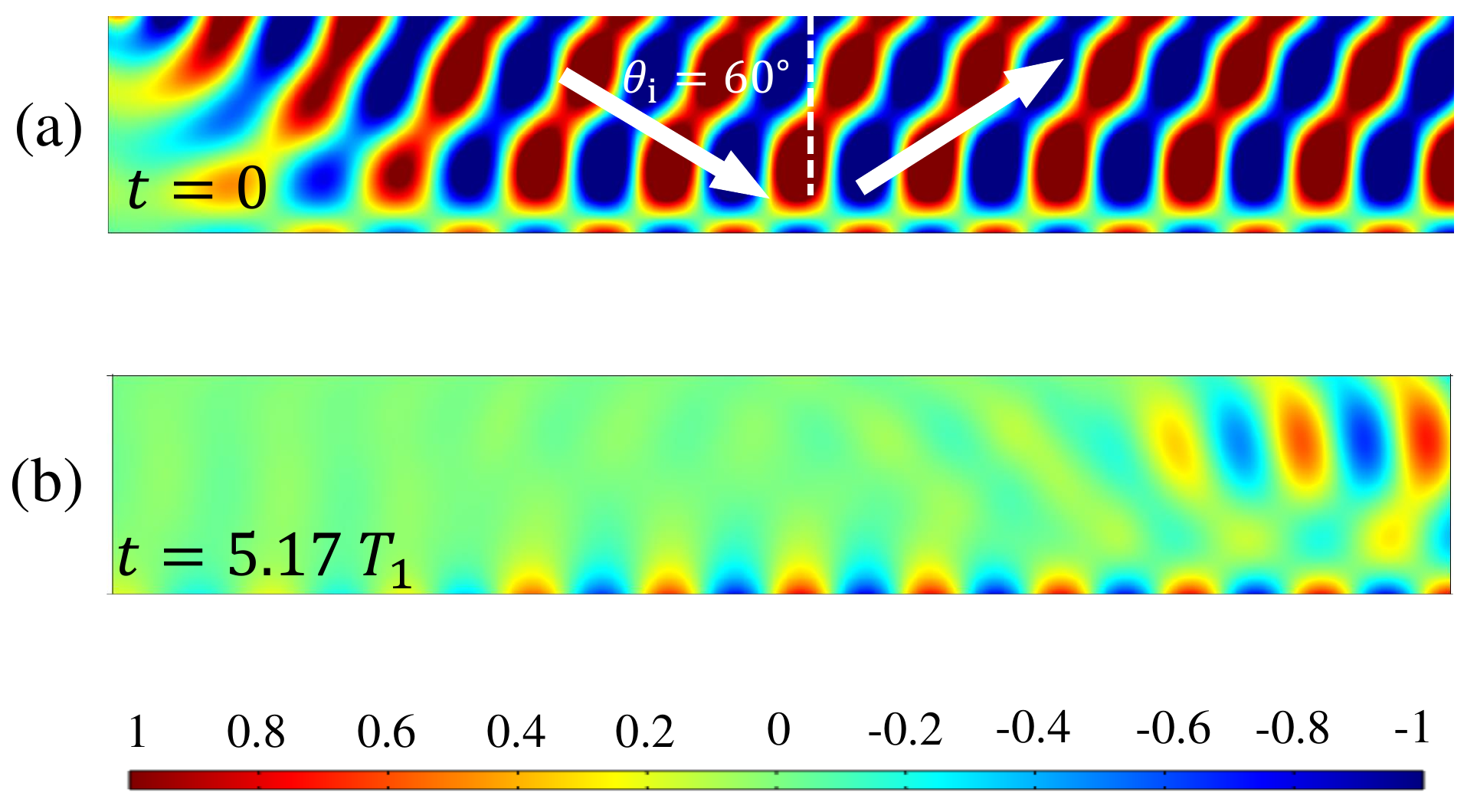}
\caption{(a) A standing plane-wave pattern is formed above the boundary $C_0$  before the temporal interface at $t=0^-$. (b) Plane waves are coupled to the surface wave after switching the capacitance to $C_1$. Field is snapped at $t=5.17T_1$.} \label{fig:p_s_conversion.pdf}
\end{figure}

It is known that surface waves cannot be excited directly by free-space propagating waves due to the mismatch of the propagation constants along the surface.  The most conventional way to couple a propagating plane wave to surface waves relies on a prism where the tangential wavevector of the evanescent wave behind the reflecting interface can be brought to phase synchronism with  the surface mode.
Here, we show that a temporal boundary can act as a temporal prism that transforms free-space propagating waves to surface waves. 
Let us consider a TE-polarized plane wave that is obliquely incident on a capacitive boundary $C_0=10$~pF, as shown in Fig.~\ref{fig:p_s_conversion.pdf}(a). Since the incident tangential wavevector $\beta=k_0\sin \theta_{\rm i}$ ($\theta_{\rm i}$ is the incidence angle and $k_0$ is the free-space wavenumber) is not on the dispersion curve of the boundary, the incident wave does not couple to the surface mode and is fully reflected in the specular direction. 
Above the boundary, there is an  interference pattern of the incident and reflected plane waves, as shown in Fig.~\ref{fig:p_s_conversion.pdf}(a). 
At $t=0$, the sheet capacitance changes rapidly from $C_0$ to $C_1=1$~pF and the excitation is removed. 
In order to adapt to the new boundary condition and excite a surface eigenmode of boundary $C_1$, the system generates transition radiation.

Depending on the capacitance value, the temporal jump creates fields at new frequencies that are  positioned on the dispersion curve of the new boundary. Therefore, the velocity of the surface wave can be controlled by the capacitance of the new boundary. If the capacitance is very large, it can couple a propagating wave to ultra-slow surface waves.

Another conventional route of coupling surface waves and plane waves in space is the use of a periodical structure. By spatially varying material properties in a periodic way, the  transverse wavevector of incident plane waves acquires a spatial momentum that allows coupling to surface modes. In an analogy, very recent works show that a temporal grating can provide a momentum that couples a plane wave to a surface wave \cite{galiffi2020wood}. This is another demonstration of the dual properties of space and time.  
 

\section{Conclusions}

To summarize, this work has studied surface waves supported by reactive metasurface boundaries that undergo sudden changes of surface properties in time. It has been shown that such temporal jumps can efficiently control the speed and amplitude of scattered surface waves. Moreover, we have found that a surface wave can be completely ``frozen'' by switching the surface reactance from capacitive to inductive, and then revived by jumping back to a capacitive boundary. In addition, we have also found that a temporal boundary can convert a free-space propagating wave to a surface wave. All these intriguing phenomena described in this paper (theoretically and numerically) explicitly indicate that how temporal modulation of electromagnetic and optical systems gives novel possibilities for controlling waves in the desired way. In future, experimental validation of these phenomena is indeed an important research direction which needs to be done.


\section*{Acknowledgments}
This work was supported  by the Academy of
Finland under Grant No. 330260.


\bibliography{references}
\bibliographystyle{IEEEtran}

\end{document}